\documentclass[aps,prl,twocolumn,footnoteinbib]{revtex4-1}

\usepackage{epsfig}
\usepackage{amssymb}
\usepackage{bm}
\usepackage{hyperref}
\usepackage{color}
\usepackage{amsmath}

\newcommand{\be}{\begin{equation}}
\newcommand{\ee}{\end{equation}}
\newcommand{\bea}{\begin{eqnarray}}
\newcommand{\eea}{\end{eqnarray}}

\newcommand{\nn}{\nonumber}

\newcommand{\anue}{\bar{\nu}_e}

\newcommand{\anum}{\bar{\nu}_\mu}

\begin{document}

\title{Meter-baseline tests of sterile neutrinos at Daya Bay}

\bigskip

\author{Y.~Gao$^{1}$ and D.~Marfatia$^{2}$\\[2ex]
\small\it $^{1}$Department of Physics, University of Oregon, Eugene, OR 97403, U.S.A.\\
\small\it $^{2}$Department of Physics and Astronomy, University of Kansas, Lawrence, KS 66045, U.S.A.}
\bigskip

\begin{abstract}
We explore the sensitivity of an experiment at the Daya Bay site, with a point radioactive source and a few meter baseline, to neutrino oscillations involving one or more eV mass sterile neutrinos. We find that within a year, the entire 3+2 and 1+3+1 parameter space preferred by global fits can be excluded at the $3\sigma$ level, and if an oscillation signal is found, the 3+1 and 3+2 scenarios can be distinguished from each other at more than the $3\sigma$ level provided one of the sterile neutrinos is lighter than 0.5~eV.
\end{abstract}

\maketitle


{\bf{Introduction.}} The standard three neutrino (3$\nu$) picture has been successful in explaining
most oscillation data. However, data
from the Liquid Scintillator Neutrino Detector (LSND) experiment~\cite{Aguilar:2001ty} when interpreted as arising from $\anum \rightarrow \anue$ oscillations,  indicate a deviation from the simple $3\nu$ picture. The Mini-Booster Neutrino Experiment (MiniBooNE)~\cite{AguilarArevalo:2008rc} provides supporting evidence for the LSND result that oscillations involving an eV mass sterile neutrino may be at work. 
Additional support may be found in an upward revision in the estimate of the reactor $\bar{\nu}_e$ flux yield~\cite{Mention:2011rk}. The fact that short baseline (SBL) reactor neutrino experiments do not detect the 3\% larger flux (via a 7\% larger event rate) could be explained as a consequence of oscillations to sterile states. 

Popular scenarios that are consistent with the the relevant data have either one sterile neutrino, with a 3+1 mass spectrum (such that the nearly degenerate triplet of mass eigenstates is lighter than the remaining state), or 2 sterile neutrinos~\cite{Kopp:2011qd, Giunti:2011gz}. The
 5 neutrino ($5\nu$) case has 2 viable spectra: a 3+2 spectrum in which the triplet is lighter than both sterile neutrinos, and a 1+3+1 spectrum in which one sterile neutrino is lighter than the triplet and one is heavier. In all cases, the sterile neutrinos mix little with the active neutrinos.

Recently, it was suggested that a ten kilocurie scale $^{144}$Ce-$^{144}$Pe $\beta$-decay source could be placed inside a large liquid scintillator 
detector to study eV sterile neutrino oscillations on baselines of a few meters with 1.8-3.3 MeV neutrinos~\cite{Cribier:2011fv}. Distinct virtues of this technique are (1) that with a point-like source, an oscillation signature can be demonstrated as a function of both energy and baseline, (2) the short baseline may be easily adjustable, (3) existing detectors can be utilized, and (4) antineutrino source activity is reduced relative to that of neutrino sources previously used for the calibration of low-energy radiochemical solar neutrino experiments since the inverse beta-decay cross section is higher than the neutrino-electron scattering cross section. Clear technical challenges are the feasibility of constructing such an intense radioactive source and of engineering suitable ultra-pure shielding of the source inside the detector.  For a decisive measurement, Ref.~\cite{Dwyer:2011xs} considered the possibility of an experiment at the Daya Bay site with a 500~kCi ($1.85\times 10^{16}$ Bq) source. The configuration of the 4 detectors in the Far Hall at Daya Bay makes it possible to place the source outside the detectors thus circumventing one of the technical issues. We treat the 500~kCi source as point-like although it will have a finite spatial extent depending on the freshness of the fuel being used for its production, the production and transportation time, as well as the final density of cerium oxide that is limited to about 4.5~g/cm$^3$. This approximation is valid since the size of the source will be small compared to the 6.5~m oscillation length of interest.

In this Letter we show that the parameter space preferred by global fits in the 3+1, 3+2 and 1+3+1 scenarios will be stringently tested by the proposed 
multi-meter baseline $\anue$ disappearance measurement at Daya Bay. For sterile neutrino masses below 0.5~eV, such a measurement can even distinguish between the 3+1 and 3+2 scenarios at the $3\sigma$ level. This enhanced sensitivity arises because knowledge of the $\nu_e$ fraction of the $\nu_4$ and $\nu_5$ mass eigenstates breaks the degeneracy in the sterile mixings to 
$\nu_e$ and $\nu_\mu$, both of which are required to explain the anomalous SBL data.


\label{sect:scenarios}

{\bf{Sterile neutrino oscillations.}} For vacuum oscillations of MeV neutrinos from a radioactive source, the (CP phase-independent) 
$\nu_e$ and $\anue$ survival probability at distance $L$ is
\be 
P_{ee}=
1- 4\sum_{i<j} \left|U_{ei} \right|^2
\left|U_{ej} \right|^2 \sin^2{\Delta_{ij}}\,,
\label{eq:surv}
\ee
where $\Delta_{ij}=\delta m^2_{ij}L/(4E_\nu)$ with $\delta m^2_{ij}=m^2_i-m^2_j$. $i,j$ denote the mass eigenstates and take values from 1 to the total number of neutrinos. $U_{ei}$ are elements of the mixing matrix. For the 3+1 spectrum, $\delta m^2_{43}\simeq  \delta m^2_{42}\simeq \delta m^2_{41}\simeq 1\ \rm{eV}^2 \gg \delta m^2_{32} \simeq \delta m^2_{31} \simeq 2.4\times 10^{-3} \rm{eV}^2 \gg \delta m^2_{21} \simeq 7.5\times 10^{-5} \rm{eV}^2$.
Then, $P_{ee}^{3+1}= 1-\sin^22\theta_s\sin^2\Delta_{41}$, with the definition, $\sin{\theta_s}\equiv U_{e4}$.

\begin{figure}[h]
\includegraphics[scale=0.45]{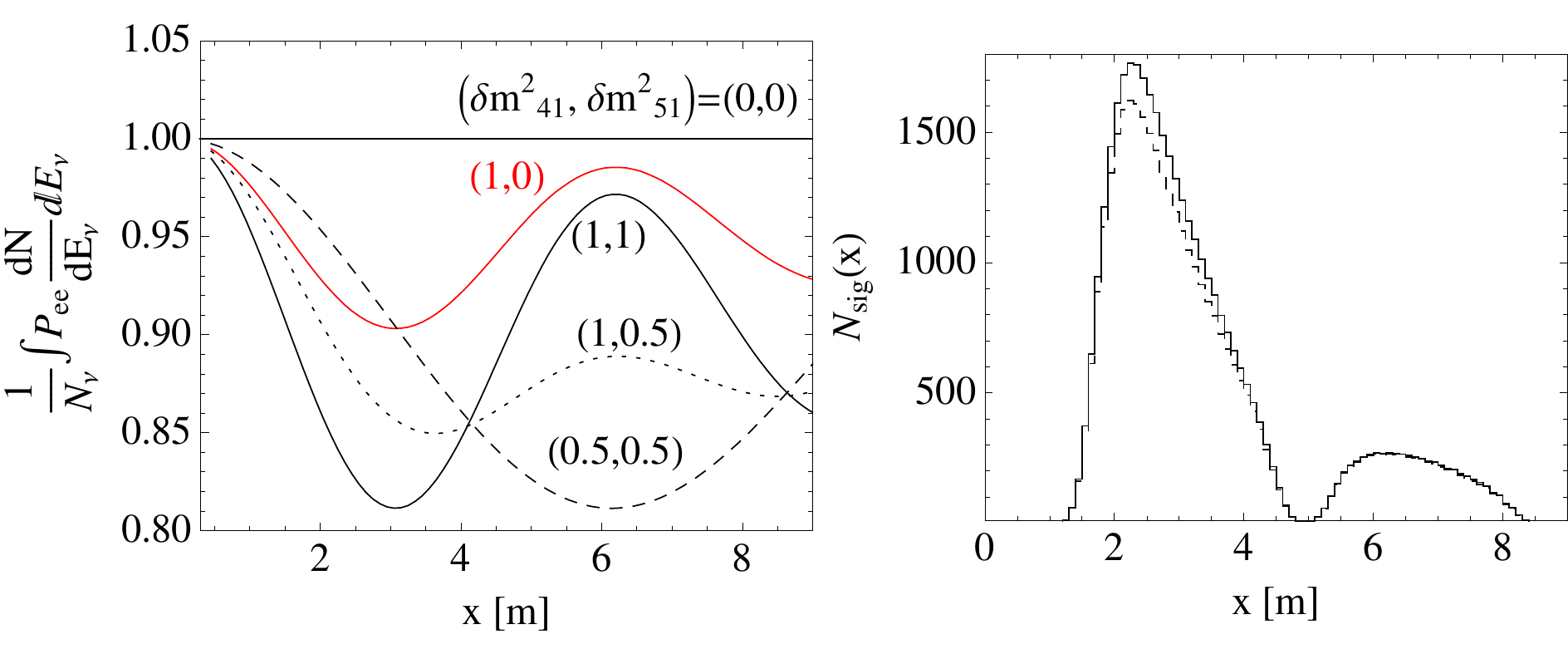}
\caption{Left: The energy-averaged $\nu_e$ survival probability as a function of distance for 3+1 and 3+2 sample points. 
$U_{e4}=0.16$ (giving a $\sim 10\%$ oscillation amplitude), and in the 3+2 scenario, $U_{e5}$ is also 0.16. 
Right: Event distributions for the chosen radioactive source-detector configuration. The solid and dashed curves show the cases of no active-sterile oscillations~\cite{Dwyer:2011xs}, and of oscillations with $\delta m^2= 1$ eV$^2$ and a 10\% oscillation amplitude, respectively. 
}
\label{fig:prob_B}
\end{figure}

In the $5\nu$ case, Eq.~(\ref{eq:surv}) includes a superposition of three oscillation frequencies corresponding to $\delta m^2_{41}, \delta m^2_{51}$ and $\delta m^2_{54}$. We neglect the $\delta m^2_{54}$ contribution in what follows.
Although the sterile neutrinos can mix with all three active neutrinos, $P_{ee}$ depends only on the four parameters, $\delta m^2_{41}, \delta m^2_{51}, |U_{e4}|$ and $|U_{e5}|$ via
\bea
P_{ee}^{5\nu}& =&
1 - 4 
(1-|U_{e4}|^2 - |U_{e5}|^2)\nn \\
&\times&(
\left|U_{e4} \right|^2 \sin^2{\Delta_{41}} 
+
\left|U_{e5} \right|^2 \sin^2{\Delta_{51}} 
)\,.
\eea
Since $P_{ee}^{5\nu}$ is insensitive to the signs of $\Delta_{41}$ and $\Delta_{51}$, $\nu_e$ disappearance
data cannot distinguish between the 3+2 and 1+3+1 spectra for identical mixing matrix elements. (In principle, the spectra can be distinguished if the suppressed but nonzero $\delta m^2_{54}$ contribution to the right hand side, $-4 |U_{e4}|^2|U_{e5}|^2\sin^2{\Delta_{54}}$, is included.)

In the left panel of Fig.~\ref{fig:prob_B}, we show the $\nu_e$ survival probability for several 3+1 and 3+2 sample points. For the sake of illustration, 
we have used somewhat large values of $U_{e4}$ and $U_{e5}$. 
The significant variation in the survival probabilities  over the first few meters for different $(\delta m^2_{41},\delta m^2_{51})$ choices reveals the strength of the method. 
For all curves in Fig.~\ref{fig:prob_B}, $P_{ee}$ is convolved with the $\anue$ energy spectrum from the radioactive source.

{\bf{Experimental set-up and procedure.}} The 500 kCi radioactive source at Daya Bay can be placed so that the 4 cylindrical detectors collect $\anue$ data with baselines from 1 to 8 meters.
Several possible source locations have been studied, each giving a different spatial coverage of $P_{ee}(L)$. 
We choose ``Point B" in the jargon of Ref.~\cite{Dwyer:2011xs}, which is located halfway between two of the detectors, and samples 2 principal baselines. It provides superior sensitivity for $\delta m^2\sim 1$~eV$^2$ with an oscillation length of about 6.5 meters. 
The no oscillation signal event rate is about 38,000 in one year after accounting for the 66.3\% decrease in source activity over a one-year 
period~\cite{Dwyer:2011xs}. Event distributions as a function of baseline are shown in the right panel of 
Fig.~\ref{fig:prob_B}; the detector energy and position resolutions are $9\%/\sqrt{E(\rm MeV)}$ and
15~cm, respectively~\cite{Dwyer:2011xs}.
Depending on the energy window used, the reactor neutrino background is expected to lie between 22,000-32,000 events per year. However, this large background can be controlled because its shape will be known. 
\label{sect:setup}
%


We take the detectors to be identical and adopt the following $\chi^2$ for our analysis~\cite{Dwyer:2011xs}:
\be 
\chi^2 = \sum_{i,j}\frac{\left(N^{ex}_{i,j}-N^{th}_{i,j}\right)^2}
{N^{ex}_{i,j}(1+\sigma_b^2 N_{i,j}^{ex})}
+\left(\frac{\alpha_s}{\sigma_s}\right)^2
+\left(\frac{\alpha_r}{\sigma_r}\right)^2\,,
\label{eq:likelihood}
\ee
where $N^{ex}_{i,j}$ is a simulated dataset and $N^{th}_{i,j}$ is the theoretical expectation for a given set of oscillation parameters, 
and $i$ and $j$ run over position and visible energy bins, respectively. 
$\sigma_s=0.01$ and $\sigma_r=0.01$ are the normalization uncertainties in
the signal and reactor background fluxes, respectively, and $\sigma_b=0.02$ is the bin-to-bin uncertainty~\cite{Dwyer:2011xs}.
$\alpha_s$ and $\alpha_r$ are nuisance parameters that are allowed to float.  
$N^{ex}$ and $N^{th}$ are given by
\be
N_{i,j}^{th/ex}= 
(1+\alpha_s)\tilde{S}_{i,j}^{th/ex} 
+(1+\alpha_r) \tilde{R}_{i,j}\,,
\label{eq:nij}
\ee
 where $\tilde{S}$ and $\tilde{R}$ (=28,000/year) are the number of signal events from the source and the number of reactor background events, respectively. 

The number of signal events (in all 4 detectors) with sterile neutrino oscillations is obtained by scaling the number of events for the $3\nu$ case:
\bea 
\tilde{S}_{i,j}^{th} &=& P_{ee}(L_i, E_\nu) S_{i,j}^{3\nu} \nn \\
{\rm{with}}\ \ S_{i,j}^{3\nu} &=& N_{tot}
\left. \frac{\Delta n}{\Delta E_{vis} } \right|_{i}
\left. \frac{\Delta n}{\Delta L           } \right|_{j}, 
\label{eq:sij}
\eea  
where $\Delta n/\Delta E_{vis}$ and $\Delta n/\Delta x$ are normalized event distributions binned in visible
energy and position, respectively, and $N_{tot}=38,000$ is the total number of events for the $3\nu$ case in one year. The positron's energy in an inverse neutron $\beta$-decay event is $E_{\nu}-(m_n-m_p)$. Subsequent pair annihilation in the scintillator produces visible energy,
\be 
E_{vis} = E_{{\nu}} -(m_n-m_p) +m_{e} \simeq E_{{\nu}} - 0.8\text{\ MeV}\,.
\ee

\begin{figure}[t]
\includegraphics[scale=0.62]{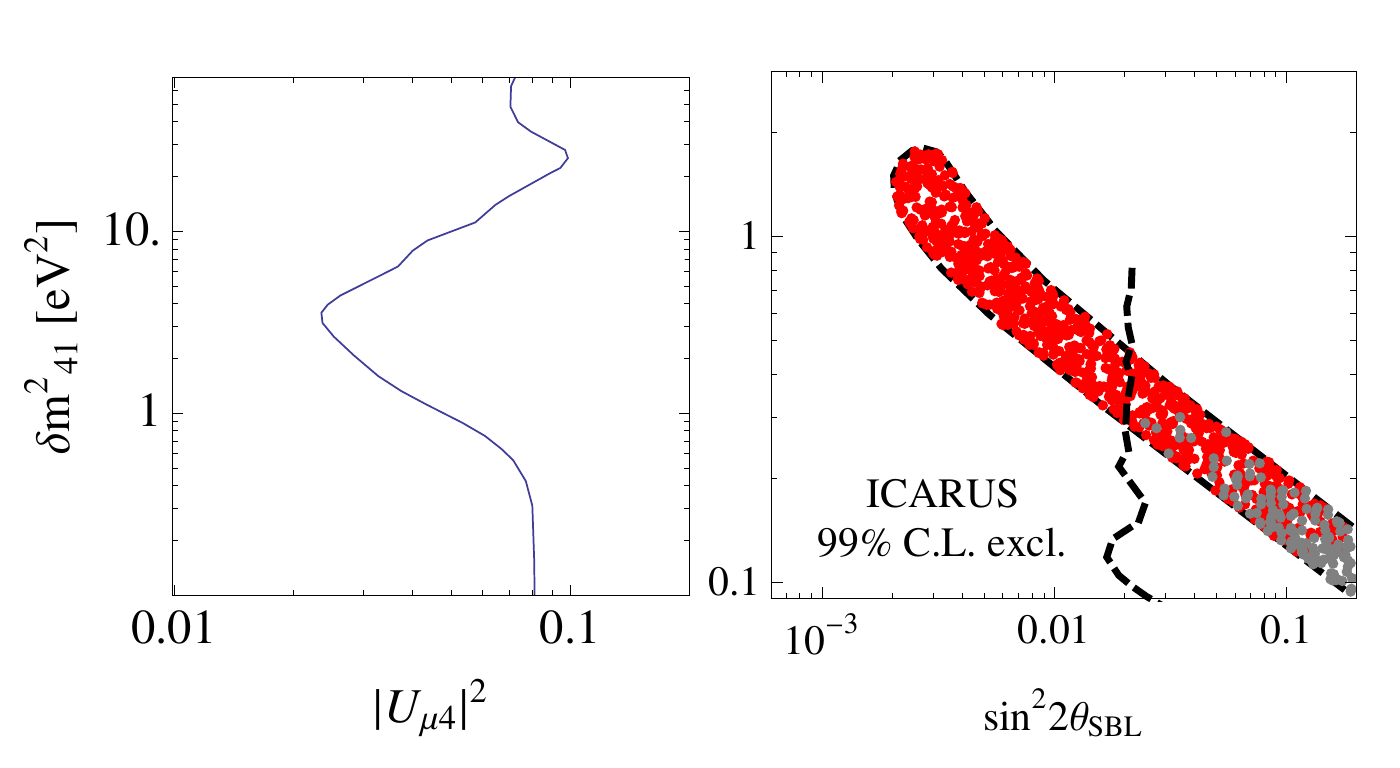}
\caption{A combination of the 99\% C.L. bound on $|U_{\mu 4}|^2$~\cite{GonzalezGarcia:2007ib} from CDHS and atmospheric neutrino data (left) and a 99\% C.L. null result at Daya Bay can rule out the red points of the 99\% C.L. region favored by a joint analysis of LSND and MiniBooNE antineutrino data in the 3+1 scenario~\cite{Kopp:2011qd} (right). The grey points survive the joint constraint, but not the ICARUS exclusion~\cite{Antonello:2012pq}.}
\label{fig:U24_bound}
\end{figure}

{\it{\bf 3+1.}}  We checked that in the 3+1 scenario our procedure yields a 95\% confidence level (C.L.) sensitivity that is comparable to that of 
Ref.~\cite{Dwyer:2011xs} for $\delta m^2_{41}<2$~eV$^2$. 
The oscillation amplitude that fits the global SBL data  is given by
\be 
\sin^2{2\theta_{SBL}} = 4|U_{e4}|^2|U_{\mu 4}|^2\,.
\label{eq:theta_sbl}
\ee
Daya Bay data could push $|U_{e4}|$ down far enough that the value of $|U_{\mu 4}|$ needed to obtain an amplitude that explains the SBL data could
conflict with the current bound on $|U_{\mu 4}|$ shown in the left panel of Fig.~\ref{fig:U24_bound}. 



Since a meter-baseline measurement at Daya Bay will be independent of the earlier data, it is reasonable to impose the constraint on $U_{\mu4}$
as a prior. Then, Daya Bay can rule out most of the allowed region from a fit to LSND and MiniBooNE antineutrino data; see the right panel of 
Fig.~\ref{fig:U24_bound}. 

\begin{figure}[t]
\includegraphics[scale=0.55]{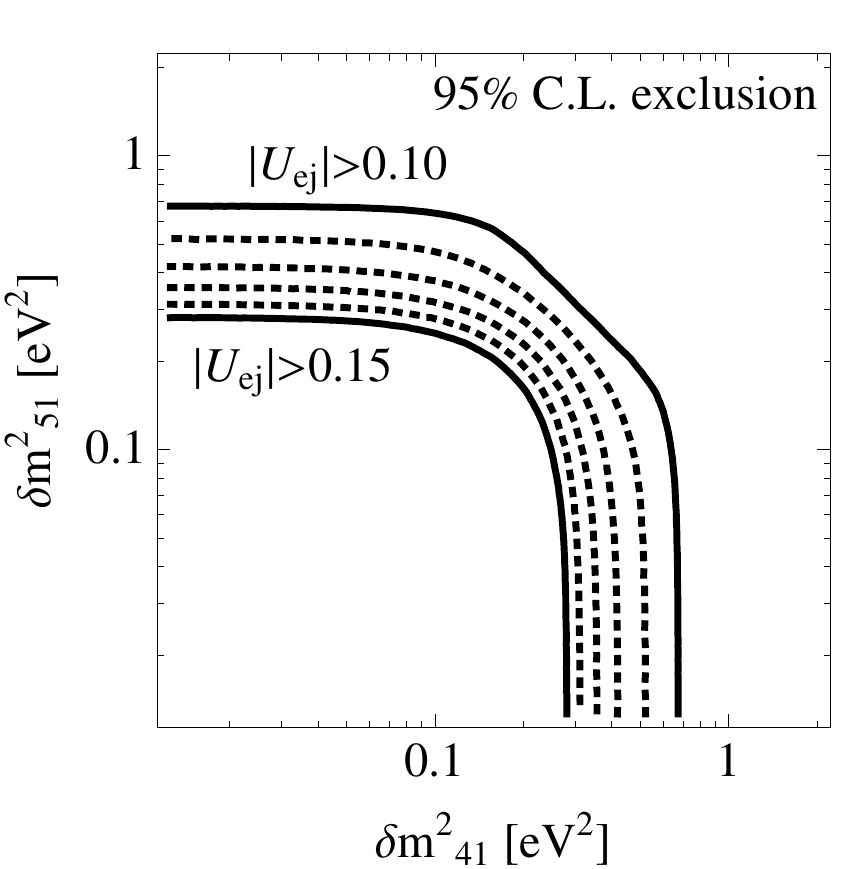}
\caption{95\% C.L. exclusion contours for the $5\nu$ scenario, for $|U_{e4}|$ and $|U_{e5}|$ above $|U|_{min}$ in steps of 0.01. The region above each contour is excluded.
}
\label{fig:2s_4par_interp}
\end{figure}

{\bf{3+2 and 1+3+1.}} 
We first consider Daya Bay's sensitivity to the $5\nu$ scenario without recourse to specific points, models or fits. 
We employ a grid in the $(\delta m^2_{41},\delta m^2_{51}, |U_{e4}|, |U_{e5}|)$ parameter space, place a prior on the size of the mixing, min$(|U_{e4}|,|U_{e5}|)=|U|_{min}$ in steps of size 0.01 from 0.10 to 0.15, and suppose a null result at Daya Bay.
The 95\% C.L. sensitivity in the $(\delta m^2_{41},\delta m^2_{51})$ plane is shown in
Fig.~\ref{fig:2s_4par_interp}.
As mentioned before, $P_{ee}$ does
not depend on the signs of the mass-squared differences. So the results of Fig.~\ref{fig:2s_4par_interp} apply to both the 3+2 and 1+3+1 spectra.

We now specialize to $5\nu$ models that are consistent with global neutrino data. In Table~\ref{tab:benchmark}, we display Daya Bay's sensitivity to several best-fit points to SBL data in the $5\nu$ case assuming that no oscillations are seen in the Daya Bay dataset. These points would be completely excluded by Daya Bay because of their sizable $U_{e4}$ and $U_{e5}$.

\begin{table}[t]
\begin{tabular}{l|l|l l|l l l l}
\hline
Parametrization&$\chi^2$ &$\delta m^2_{41}$& $\delta m^2_{51}$ & $|U_{e4}|$ & $|U_{e5}|$ & $|U_{\mu 4}|$ & $|U_{\mu 5}|$ \\
\hline
KMS (3+2) \cite{Kopp:2011qd}&62 &0.47 &0.87 &0.128 &0.138 &0.165 &0.148 \\
KMS (1+3+1) \cite{Kopp:2011qd}&68 & -0.47 & 0.87& 0.129& 0.142& 0.154& 0.163\\
GL (3+2) \cite{Giunti:2011gz}&78 & 0.9 & 1.61& 0.13& 0.13& 0.14& 0.078\\
MM NH (3+2)\cite{Donini:2012tt}&64 & 0.47& 0.87&0.149 &0.127 &0.112 &0.127 \\
MM IH (3+2)\cite{Donini:2012tt}&80 & 0.9 & 1.61 & 0.139 &0.122 &0.138 &0.107 \\
MMS (3+2) \cite{Fan:2012ca}&55 &0.89 &1.76 &0.15 &0.07& 0.15 &0.15 \\
\hline
\end{tabular}
\caption{$\chi^2$ values for some global best-fit points (to data from SBL experiments), for a simulated dataset
with no oscillations at Daya Bay. `MM NH/IH' is the `minimal model (with normal/inverted $3\nu$ mass hierarchy)' of Ref.~\cite{Donini:2012tt} and `MMS' is the `minimal seesaw model' of Ref.~\cite{Fan:2012ca}.}
\label{tab:benchmark}
\end{table}

\medskip

\begin{figure}[t]
\includegraphics[scale=0.58]{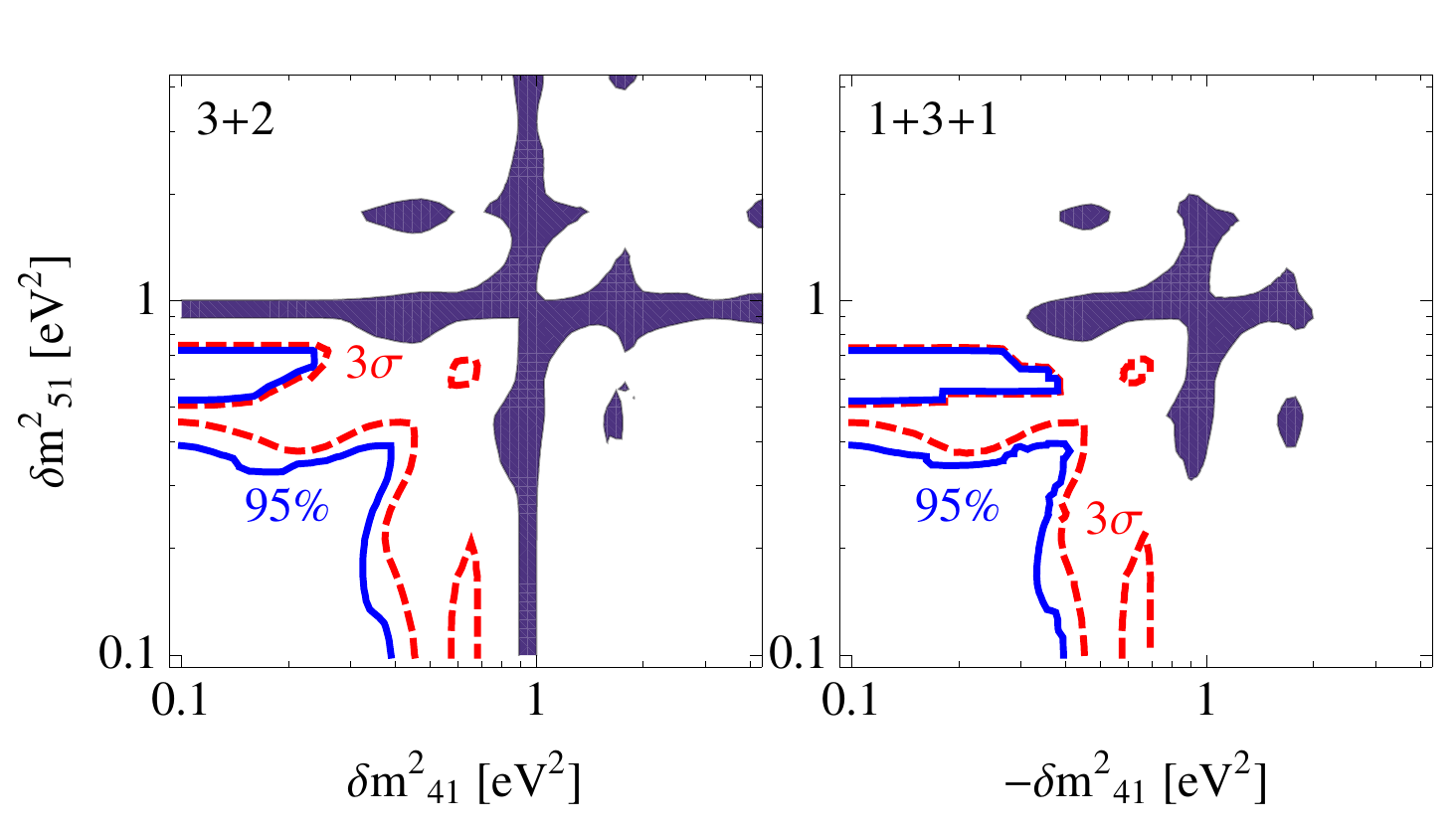}
\caption{The shaded regions are the $3\sigma$ globally preferred regions in the $(\delta m^2_{41},\delta m^2_{51})$ plane in the 3+2 (left) and 1+3+1 (right) scenarios~\cite{kopp}. A null result at Daya Bay can rule out the regions above the blue solid and red dashed contours at the 95\% C.L. and $3\sigma$, respectively. The exclusion regions are slightly different in the two panels because for each $(\delta m^2_{41},\delta m^2_{51})$, the best-fit mixing matrix elements from the fit of Ref.~\cite{kopp} are different for the two mass spectra.
}
\label{fig:3+2contour}
\end{figure}

To examine Daya Bay's capability to probe the large $5\nu$ parameter space, we use the globally allowed regions from an updated fit to the datasets 
 listed in Ref.~\cite{Kopp:2011qd} in conjunction with data from the NOMAD~\cite{Astier:2003gs} and CDHS~\cite{Dydak:1983zq}
  experiments~\cite{kopp}. The shaded areas of Fig.~\ref{fig:3+2contour} are the globally allowed regions at $3\sigma$.
  We see that at least one 
$\delta m^2$ is close to 1~eV$^2$ so as to explain the SBL data.
All mixing parameters other than $\delta m^2_{41}$ and $\delta m^2_{51}$ are marginalized over and assume their best-fit values. 

As the global fits favor significant $\anum - \anue$ transitions, the mixing parameters tend to be large enough to be testable at Daya Bay. 
Figure~\ref{fig:3+2contour} shows that Daya Bay can exclude the 3+2 and 1+3+1 scenarios as an explanation of the LSND/MiniBooNE anomaly at $3\sigma$. 


{\bf {3+1 or 3+2?}} 
\label{sect:1or2}
So far we have demonstrated that a null result at Daya Bay can significantly constrain sterile neutrinos. We now entertain the possibility that future data confirms their existence. Then, a pressing issue will be to ascertain whether the  3+1 or the 3+2 scenario is operative. 
Since scenarios with more eigenstates should be able to mimic those with fewer eigenstates, a good test of Daya Bay's discriminatory power is to fit 3+2 points to data simulated for 3+1. Assume that Daya Bay collects a dataset that is well-described by a point in the 3+1 parameter space.
Then, in principle there is a 3+2 mixing scenario that gives the same oscillation pattern. However, this 3+2 point may be constrained by other oscillation data. To account for this possibility, we fit all the globally allowed 3+2 parameters
to the 3+1 dataset and check if a good fit exists. The technical
procedure is as follows.

For every point on a grid in the $(\delta m^2_{41},\theta_{SBL}, |U_{e4}|)$ parameter space that lies 
within the 99\% C.L. allowed region of the right panel of 
Fig.~\ref{fig:U24_bound} and is also consistent with the 99\% C.L. bound on $|U_{\mu 4}|$ in the left panel of  Fig.~\ref{fig:U24_bound}, we simulate 
a dataset ${N}^{ex}_{i,j}$. 
We then fit points in 3+2 parameter space that are allowed at $3\sigma$ (shown in the left panel of Fig.~\ref{fig:3+2contour}) to this dataset 
(using Eq.~\ref{eq:likelihood}), and find the 3+2 point with the minimum
${\chi}^2$ corresponding to that $(\delta m^2_{41},\theta_{SBL}, |U_{e4}|)$ point. For a given $\delta m^2_{41}$, we repeat the procedure for other values of $(\theta_{SBL}, |U_{e4}|)$ so as to find the global $\chi^2_{min}$ for each $\delta m^2_{41}$. Note that the best-fit 3+2 value of 
$\delta m^2_{41}$ need not be the same as the value for which 3+1 data was simulated.

\begin{figure}[t]
\includegraphics[scale=0.55]{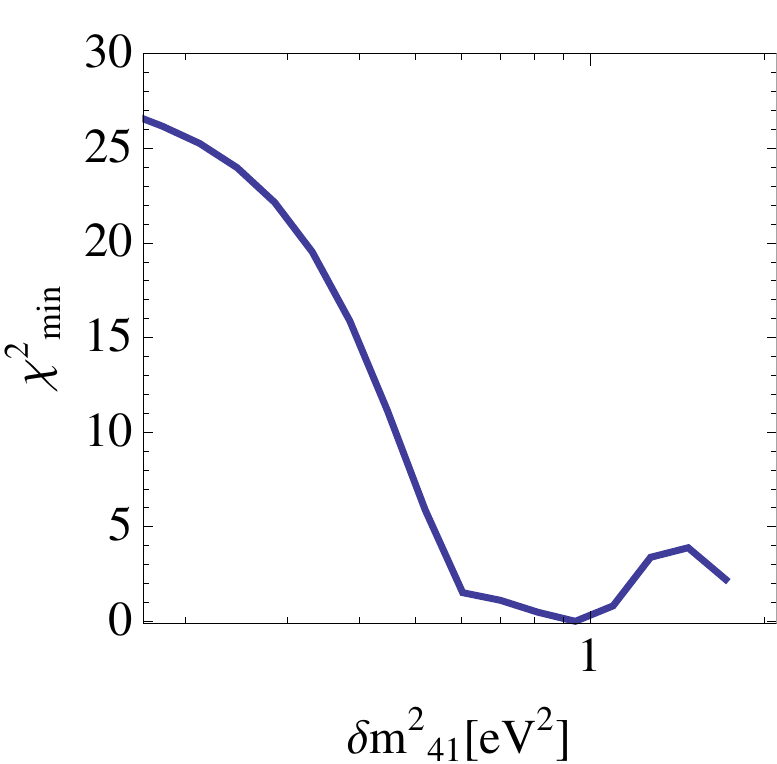}
\caption{The degree to which Daya Bay can discriminate between the 3+1 and 3+2 scenarios. We simulate an oscillation signal for points in the 99\% C.L. region favored by LSND and MiniBooNE that are consistent with the 99\% C.L. bound on $|U_{\mu 4}|$ (see Fig. ~\ref{fig:U24_bound}), and fit the spectrum from points in the $3\sigma$ region of the 3+2 parameter space (see the left panel of Fig.~\ref{fig:3+2contour}) to the simulated data. A more than $3\sigma$ discrimination is possible for $\delta m^2_{41} < 0.5$~eV$^2$. 
}
\label{fig:3+2OR3+1}
\end{figure}

We plot $\chi^2_{min}$ versus $\delta m^2_{41}$ in Fig.~\ref{fig:3+2OR3+1}. The discrimination between the 3+1 and 3+2 scenarios is better for small 
$\delta m^2_{41}$. This is because for small $\delta m^2_{41}$, the deviation of the 3+1 spectrum from the $3\nu$ spectrum is small in the meter-baseline experiment, which is harder to replicate with a 3+2 point that must also reproduce the anomalous SBL data.  

\vskip 0.1in
{\it Acknowledgements.}~
We thank V. Barger, K. Heeger and B. Littlejohn for discussions, and especially thank J.~Kopp for providing us with data from the update to the fit of
Ref.~\cite{Kopp:2011qd}. This work was supported by DOE Grants No. DE-FG02-04ER41308 and DE-FG02-96ER40969.


\end{document}